\documentclass[amssymb,prl,twocolumn,showpacs]{revtex4}
\usepackage{epsfig}
\usepackage{dcolumn}
\usepackage{amsmath}
\begin{document}

\title{Magnetic-field asymmetry of nonlinear mesoscopic transport} 
\author{David S\'anchez}
\author{Markus B\"uttiker}
\affiliation{D\'epartement de Physique Th\'eorique,
Universit\'e de Gen\`eve, CH-1211 Gen\`eve 4, Switzerland}
\date{\today}

\begin{abstract}
We investigate departures of the Onsager relations in the nonlinear regime
of electronic transport through mesoscopic systems.
We show that the nonlinear current-voltage characteristic is not an even function
of the magnetic field due only to the magnetic-field dependence
of the screening potential within the conductor.
We illustrate this result for two
types of conductors: A quantum Hall bar with an 
antidot and a chaotic cavity connected to quantum point contacts. 
For the chaotic cavity we obtain through random matrix theory an asymmetry
in the fluctuations of the nonlinear conductance that vanishes rapidly 
with the size of the contacts. 
\end{abstract}

\maketitle

{\em Introduction}.---The Onsager-Casimir relations~\cite{ons31,cas45}
are symmetry conditions for
correlation functions. These can be cast as friction
coefficients by means of the fluctuation-dissipation theorem.
In electronic transport measurements,
the information about dissipation is carried by the
conductance of the sample.
Microscopic reversibility thus requires that in the presence
of a magnetic field $B$ the conductance obeys
$G_{\alpha\beta}(B)=G_{\beta\alpha}(-B)$ between
contacts $\alpha$ and $\beta$. In particular, for a two-probe 
conductor the conductance is an even function of magnetic 
field $G (B)=G (-B)$. 
Such relations generally hold for {\em macroscopic}
systems near thermodynamic {\em equilibrium}.
What happens when these conditions are not fulfilled?
First, when transport is phase coherent as it occurs in {\em mesoscopic}
conductors, the conductance is not just material specific but
also depends on the probe configuration.
Then, a generalized reciprocity theorem may be proved provided
current and voltage probes are treated on equal footing~\cite{but86}.
Second, away from equilibrium (e.g., in {\em nonlinear} dc-transport)
there are no fundamental reasons why the Onsager relations should hold.
In particular, in a two terminal conductor, we can expect 
that the nonlinear $I$--$V$ characteristic is not an even function 
of magnetic field, $I(B,V) \ne I(-B,V)$. Below, we demonstrate 
that such asymmetries can indeed be found but importantly 
only to the extent that the potential landscape in the interior 
of the conductor becomes an uneven function of magnetic field~\cite{but93}. 

Nonlinear transport in mesoscopic conductors
is a subject of growing interest.
Rectification effects in asymmetric microjunctions~\cite{son98},
quantum point contacts~\cite{sho01}, and 
switching and gain in
three-terminal ballistic branches~\cite{rei02}
have been observed.
For samples with special spatial symmetries, 
a set of symmetry relations which hold 
in the nonlinear regime have been discussed and experimentally 
verified by L\"ofgren {\it et al.}~\cite{lof04}
Conductance fluctuations away from equilibrium have long been 
of theoretical~\cite{alt86}
and experimental~\cite{web88,tab94} concern.
Recently, Zumb\"uhl, Marcus {\it et al.}~\cite{unpub} 
started experiments to investigate the magnetic-field symmetry
of fluctuations away from equilibrium. 

A correct treatment of this problem requires a {\em self-consistent} 
discussion of the screening potential within 
the conductor. A scattering picture
of weakly nonlinear transport was provided
by Christen and B\"uttiker~\cite{but93,chr96}. 
The work of Sheng {\it et al.}~\cite{she99} exemplifies a computational 
implementation. We briefly now discuss the 
essential elements of this approach.

Consider a mesoscopic conductor penetrated
by a magnetic field $B$ perpendicular to the plane of the sample 
and connected to $\alpha=1,\ldots , M$ reservoirs and gates. 
Transport is described by
the scattering matrix $s_{\alpha\beta}$.
For purely elastic transport, the scattering matrix is a function 
of the energy $E$ of the carriers and it is a functional of the 
potential landscape $U(\vec{r})$ of the conductor, 
$s_{\alpha\beta}[E;U(\vec{r})]$. For linear transport, 
the scattering matrix is evaluated at the equilibrium potential 
$U_{\rm eq} (\vec{r})$ which is an even function of $B$. 
As a consequence,
$s_{\alpha\beta}(B) =s_{\beta\alpha}(-B)$ at equilibrium.
Away from equilibrium the potential depends on the voltages ${V_\alpha}$ 
(measured from an equilibrium
chemical potential $\mu_0$) applied
to the leads and the nearby gates. We can write 
$U(\vec{r}) = U_{\rm eq} (\vec{r}) + 
\sum_{\alpha} u_{\alpha} V_{\alpha} + {\cal O}(V^{2})$.
Here, the characteristic potentials (CP's) $u_\gamma(\vec{r})=
[\partial U(\vec{r})/\partial V_{\gamma}]_{\rm eq}$ relate the variation
of $U(\vec{r})$ in the sample
to a voltage shift in the contact $\gamma$~\cite{but93}.
We expand the current through lead $\alpha$ in powers of the voltage shifts
up to second order:
\begin{equation}
I_\alpha=\sum_\beta G_{\alpha \beta} V_{\beta} +
\sum_{\beta\gamma} G_{\alpha\beta\gamma} V_{\beta} V_{\gamma} \,.
\end{equation}
The linear conductances are expressed by the well known
formula $G_{\alpha \beta}=(e^2/h) \int dE [-\partial_E f(E-\mu_0-V_\alpha)]
A_{\alpha\beta}(E;\{V_\alpha\}=0)$,
where $f(E)$ is the Fermi function
and $A_{\alpha\beta}(E;\{V_\alpha\})= 
{\rm Tr}[\mathrm{1}_{\alpha\beta}\delta_{\alpha\beta}-
s_{\alpha\beta}^\dagger  s_{\alpha\beta} ]$.
The second-order nonlinearity $G_{\alpha\beta\gamma}$
contains information about the charge {\em response} of the system.
Reference~\cite{chr96} finds:
\begin{equation} \label{eq_gnonl}
G_{\alpha\beta\gamma}=\frac{e^2}{h} \int dE \,
\frac{\partial f}{\partial E}
\int d^3 r \, \frac{\delta A_{\alpha\beta}}{\delta U(\vec{r})}
[\delta_{\beta\gamma}-2 u_\gamma (\vec{r})]\,.
\end{equation}
Here the nonequilibrium state is described by the CP's, 
$u_{\beta}$, which arise as a consequence of screening 
of the additional bare charge injected from contact $\beta$.
This charge density of states (DOS)
is the injectivity \cite{but93} of contact $\beta$,
\begin{equation} \label{eq_inject}
\frac{d\bar{n}_{\beta}(\vec{r})}{dE} = -\frac{1}{2\pi i} \sum_{\alpha}
{\rm Tr}\left(s^{\dagger}_{\alpha\beta}
\frac{\delta s_{\alpha\beta}}{e\delta U(\vec{r})}\right)\,.
\end{equation}
On the other hand, the additional current in contact $\beta$ 
due to a variation 
of the screening potential at point $\vec{r}$,
is given by the emissivity into $\beta$, 
$d\underline{n}_{\beta}(\vec{r})/dE$.
Due to microreversibility, $d\bar{n}_{\beta}(B,\vec{r})/dE=
d\underline{n}_{\beta}(-B,\vec{r})/dE$.
However, neither the injectivity nor the emissivity alone
are invariant under $B$-reversal.
As a result, the CP's are {\em not} even functions of $B$.
Thus, quite generally, even in a two-terminal setup 
the second-order contribution $G_{\alpha\beta\gamma}$
to the $I$--$V$ characteristic is not even in the field.
We define the magnetic-field asymmetry for such a setup:
\begin{equation}\label{eq_as}
\Phi = \frac{1}{2} [G_{111}(B)-G_{111}(-B)]\,.
\end{equation}
We emphasize that the asymmetry $\Phi\neq 0$ is generated
in Eq.~(\ref{eq_gnonl})
only through the asymmetry in the electric potential: if the 
potential is even in $B$, Eq. (\ref{eq_gnonl}) predicts for a two-terminal conductor 
a current that is even.
Thus, a self-consistent description of charge redistribution is crucial.
 
Our purpose is to elucidate
this general result with the help of a simple but instructive example 
(a quantum Hall bar with an antidot) and 
to provide a prediction of the size of $\Phi$ for a generic conductor, 
a chaotic cavity.  
\begin{figure}[b]
\centerline{
\epsfig{file=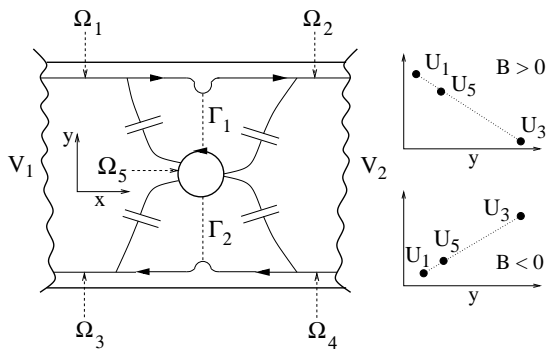,angle=0,width=0.40\textwidth,clip}
}
\caption{Left panel: quantum Hall conductor attached to two reservoirs
with an antidot connecting two edge states.
Right panel: Schematic spatial variation of the screening potential
inside the bar for opposite polarities of the magnetic field.
}
\label{fig1}
\end{figure}

{\em Quantum Hall bar}.---We consider a
conductor in the quantum Hall regime as depicted
in the left panel of Fig.~\ref{fig1}. Edge states are symbolically indicated 
by arrows along the upper and lower edge of the sample. 
For simplicity, we assume that $B$ is so strong that only the lowest Landau
level is occupied (filling factor $\nu=1$).
Backscattering is achieved by
producing with gates a potential hill,
thus forming an antidot~\cite{for94}.
The antidot behaves effectively
like a quantum impurity with a Breit-Wigner 
resonance at $E_0 + eU_d$ coupled to the edge states via
hybridization widths $\Gamma_1$ and $\Gamma_2$ due to tunneling.
Both broadenings are taken to be energy independent. 
$U_d$ is the potential at the antidot. 

Let us first provide an intuitive picture of the nonlinear transport.
Application of a voltage $V_1>V_2$, leads to an excess charge 
on the upper edge, and through screening to a corresponding deficit 
of charge on the lower edge.
Hence, a Hall potential $U_H (B,\vec{r})$ is established.
The charges on the edge state lifts the resonant energy 
to a value $E_0 + eU_d (B)$ (upper right panel of Fig.~\ref{fig1}). 
If we now reverse the field, it is the lower edge state that is charged, 
and the upper edge state that is lowered in energy through screening, 
generating an electric field opposite to that of $U_H (B)$. 
Then, the resonant energy will now be at $E_0 + eU_d (-B)$
(lower right panel of Fig.~\ref{fig1}). 
Clearly, $eU_d (B)$ and $eU_d (-B)$ will in general not be equal,
except if stringent symmetry conditions are fulfilled. 
The symmetry is broken either through {\em scattering asymmetry}, i.e.,
if transmission into the antidot is not symmetric $\Gamma_1 \ne \Gamma_2$, or 
through {\em electrical asymmetry}
if the charges on the upper
edge couple more strongly to the antidot than the
charges on the lower edge.

We now support this picture with analytical calculations. 
Evaluation of the exact local potential distribution
can typically be achieved only computationally. 
To simplify the problem, we divide the conductor into 
five regions $\Omega_i$ with $i=1,\ldots , 5$ as indicated in Fig.~\ref{fig1}. 
$\Omega_5 \equiv \Omega_d$ is the region of the edge state circling the 
antidot. In each region the 
potential is taken to be constant. Interaction between charges in different 
regions is described by a geometric capacitance matrix $C_{ij}$. 
Such a {\em discrete} potential model captures the 
essential physics~\cite{but93,chr96}. Thus, the CP
in region $i$ is $u_{i\gamma}= \partial U_i/\partial V_\gamma$
and the injectivity of lead $\alpha$ into region $i$ is
$\bar{D}_{i \alpha}=\int_{\Omega_i} d^3r \, d\bar{n}_{\alpha}(\vec{r})/dE$.
When the energy is close to $E_0 +eU_d$ an electron is reflected
through the antidot (e.g., from $\Omega_1$ to $\Omega_3$) with
a probability $R=1-T=\Gamma_1\Gamma_2/|\Delta|^2$, where
$\Delta=\mu_0-E_0-eU_5+i\Gamma /2 $ with $\Gamma=\Gamma_1+\Gamma_2$
as the total linewidth. 
Using the corresponding scattering matrix
we find the injectivities into the different regions. 
For the case $B>0$ (Fig.~\ref{fig1}, left panel) the DOS
associated to carriers at $\Omega_3$ injected by lead 1 reads 
$\bar{D}_{31}=D_3 R$, where $D_i$ is the total
($B$-dependent) DOS of the upper edge state in region $i$.
Similarly, the remaining injectivities of contact $1$
are $\bar{D}_{11}=D_1$, $\bar{D}_{21}=D_2 T$, $\bar{D}_{41}=0$ and 
$\bar{D}_{51}=e^2\Gamma_1/2\pi|\Delta|^2$.
One proceeds likewise to find the injectivities of contact $2$,
$\bar{D}_{i 2}$. The charge $q_i$ in region $i$ can be expressed in two ways:
\begin{equation}
\label{eq_q}
q_i = e^{2} \sum_{\alpha} \bar{D}_{i\alpha }
(V_\alpha - U_i) = \sum_j C_{ij} U_j \,. 
\end{equation}
First, $q_i$ is the bare injected charge due to the voltage 
applied to the contact and the screening charge
induced by the internal potential $U_i$.
Second, $q_i$ is the charge permitted by the Coulomb interaction, 
where $C_{ij}$ is the geometrical capacitance matrix whose indices run 
over all (five) regions considered in Fig.~\ref{fig1}.
Equation~(\ref{eq_q}) allows us to determine the 
potentials $U_i$ as a function 
of the applied voltages. 
We take equal DOS for all regions $D_i=D$.
We consider separately the case of (A) an electrically symmetric sample
that is asymmetric only in the scattering properties and (B) 
the case of a sample with symmetric scattering that is asymmetric 
only electrically. 

For case (A) we assume a capacitance matrix
with equal capacitances $C$ between the edge  
states and the antidot.
Transmission from the antidot to the upper and lower edge state is asymmetric 
$\eta=(\Gamma_1-\Gamma_2)/\Gamma$.
Using Eq. (\ref{eq_gnonl}) and Eq. (\ref{eq_q}) we evaluate 
the second-order nonlinear conductance and insert it in Eq. (\ref{eq_as}).
Taking into account 
that the transmission (reflection) probabilities depend only on the 
potential at the antidot, and that ${\delta T}/e{\delta U_d} = -dT/dE|_{E_F}$
at zero temperature, we find 
\begin{equation}
\label{scattasym}
\Phi = -\frac{e^3}{h}
\left. \frac{dT}{dE}\right|_{\rm eq}
\frac{\eta R(C+ e^{2}D )}{2\pi CD \Gamma+R(C+ e^{2}D )}
+\mathcal{O}(\eta^3)\,.
\end{equation}
We observe that in the charge neutral limit ($C=0$) the magnetic-field
asymmetry $\Phi$ is independent of $R$ (to leading order in $\eta$).
In the noninteracting limit  ($C\to \infty$), the asymmetry
is proportional to $R/(R+2\pi D\Gamma)$. 

Next consider case (B), a sample with symmetric
scattering properties ($\eta =0$)
but being electrically asymmetric. Such a case arises if 
the tunnel barrier separating the antidot from the top edge state is 
not very high in energy but wide whereas the barrier separating the 
antidot from the lower edge state is high in energy but narrow such that 
transmission through both is equal. However, the capacitance 
to the upper $(1+\xi)C$ and lower $(1-\xi)C$ edge states will differ, 
$\xi$ being a dimensionless parameter.
As above, we calculate the $B$-dependent second-order conductance
to lowest order in $\xi$ to find: 
\begin{equation}
\Phi =-\frac{e^3}{h}
\left. \frac{dT}{dE}\right|_{\rm eq}
\frac{\pi \xi e^{2}D^{2}C\Gamma T}{(C+ e^{2}D)[2\pi CD\Gamma+R(C+e^{2}D)] }\,.
\end{equation}
Interestingly, the magnetic-field asymmetry is proportional
to the transmission probability $T$, unlike case (A). 
Thus, the transmission of the impurity would act experimentally
as the indicator of the physical mechanism behind
the resulting field asymmetry.

Our discussion demonstrates that either asymmetric scattering
or an electrical asymmetry generates already to second order in voltage 
a deviation from the Onsager relations that hold in the linear regime. 
For strong backscattering, this deviation
is due mainly to a scattering induced asymmetry whereas
for weakly coupled impurities the electric asymmetry dominates.
\begin{figure}[b]
\centerline{
\epsfig{file=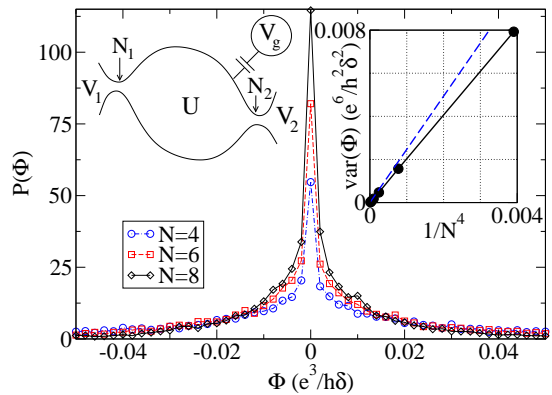,angle=0,width=0.40\textwidth,clip}
}
\caption{(Color online). Normalized probability densities for the magnetic-field
asymmetry $\Phi$ of a chaotic cavity connected to two reservoirs
(left inset). Right inset: fluctuations of $\Phi$ as a function
of the total number of modes $N$. The dashed line is the analytical prediction
for $N\gg 1$.
}
\label{fig2}
\end{figure}

{\em Chaotic cavity}.---It is important to find out whether such symmetry
breaking is only 
relevant for few-channel problems or whether in fact symmetry 
breaking is also observable in the many-channel limit. This is the motivation
to investigate now a chaotic cavity, which is
a metallic quantum dot
whose classical analog displays chaotic dynamics.
For clean samples the transport
within the cavity is ballistic and its corresponding statistics
is well described by random matrix theory~\cite{bar94}.
Open cavities have been extensively studied both theoretically~\cite{bee97}
and experimentally~\cite{mar97}. Most of these works are restricted
to the linear conductance regime (see, however, Ref.~\cite{lof04}).  
The chaotic cavity is coupled to reservoirs ($\alpha=1,2$) through
quantum point contacts with $N_1$ and $N_2$ propagating channels
(left inset of Fig.~\ref{fig2}). On the ensemble average 
such a cavity exhibits simply a linear $I$--$V$ characteristic with a conductance 
$G = e^{2} N_1 N_2 /hN$ where $N=N_1 + N_2$.
Nonlinearities arise due to quantum 
fluctuations with an energy scale equal to the Thouless energy
$E_T= N\delta$ with $\delta$ the mean level spacing. 

Since a cavity is effectively zero-dimensional due to its isotropic properties, 
we take into account screening with a single potential $U$.
Surrounding gates are coupled capacitively with a gate voltage $V_g$
and a geometric capacitance $C$. In response to a shift of the
contact voltages $V_\alpha$
a nonequilibrium charge builds up in the cavity which depends on the 
injecting contact. The two equations which determine the excess charge 
on the cavity are 
\begin{equation}\label{eq_cav}
Q = \sum_{\alpha} e^2 \bar{D}_\alpha (V_\alpha - U)
= C (U- V_g) \,,
\end{equation}
where, as before, $\bar{D}_\alpha$ is the injectivity of lead $\alpha$.
The total DOS of the cavity is $D=\sum_{\alpha} \bar{D}_\alpha$.

As we have now a three-terminal problem with a gate voltage 
$V_3 = V_g$, we consider the case where $V_3 = V_2$.
From Eq. (\ref{eq_gnonl})
we get $I_1 = G_{11}(V_1 -V_2) + G_{111} (V_1 - V_2)^{2} +{\cal O}(V^3)$.
Using the WKB approximation, 
we can again replace derivatives with regard to potentials by 
energy derivatives~\cite{chr96} and find   
\begin{equation}\label{eq_g111}
G_{111}= - \frac{e^3}{h} \left.\frac{dT}{dE}\right|_{\rm eq}
(1-2 u_1)\,.
\end{equation}
For the chaotic cavity the ensemble average 
$\langle G_{111}\rangle=0$ vanishes, 
an asymmetry can develop only due to quantum {\em fluctuations}.  
We now show that ${\rm var} (\Phi)$ is nonzero.

For $N_1\gg 1$ and $N_2\gg 1$ within random matrix theory 
the product in the right hand side of Eq. (\ref{eq_g111}) can be
decoupled~\cite{bee97}. Thus, we disregard correlations
between $dT/dE$ and $u_1$.
The {\em unscreened} nonlinear conductance
$- (e^3/h) dT/dE|_{\rm eq}$
changes sign randomly on the ensemble so that its average is zero~\cite{pol03}.
In the unitary ensemble (magnetic flux through the cavity of the 
order of $1$ quantum) 
we find for the fluctuations ${\rm var} (dT/dE) = 8\pi^2 N_1^2 N_2^2/N^6 \delta^{2}$.
Furthermore, we neglect the small fluctuations~\cite{bro97} in the
electrochemical capacitance
$1/C_{\mu} = 1/C + \delta/e^{2}$.
Then, from Eq.~(\ref{eq_cav}) and Eq.~(\ref{eq_g111})
we find that the fluctuations of $u_1$
are determined by the fluctuations of the asymmetric part
of the injectivity. Correlations of injectivities 
have been investigated in Ref.~\cite{bro97}. Using these results
in Eq. (\ref{eq_as}) we find 
\begin{equation}
\label{eq_varphi}
{\rm var}(\Phi)=\frac{e^6}{h^2\delta^2}\frac{16 \pi^2 N_1^3 N_2^3}{N^{10}}
\left( \frac{C_{\mu}}{C} \right)^2\,.
\end{equation}
The fluctuations go as $1/N^4$ for $N\gg 1$, thus
vanishing quickly with increasing number of channels and are maximal 
for perfect screening $C_{\mu} = C$.  

For small $N_1$ and $N_2$
a full analytical calculation of ${\rm var}(\Phi)$
is involved since one has to evaluate
high-order correlations between $D_{\alpha}$, $D$,
and $dT/dE$. Yet, we can substantiate
Eq.~(\ref{eq_varphi}) with numerical calculations.
The results for $C=0$ and $N_1=N_2=N/2$
are shown in Fig.~\ref{fig2}. In these simulations, the
$S$ matrix is expressed in terms of the Hamiltonian matrix
of the cavity, whose elements are random from the unitary ensemble~\cite{bee97}.
From the resulting $S$ and $dS/dE$ we compute the distribution
of $\Phi$. We observe that the probability density of $\Phi$
is a narrow peak centered around zero. The variance decreases
quickly with $N$ (see the right inset).
Strikingly, the fluctuations
are in good agreement (within the numerical error)
with the analytical prediction~(\ref{eq_varphi}).
These results demonstrate
unambiguously that the fluctuations of the nonlinear
conductance of a chaotic cavity are not symmetric
under field reversal and that they depend strongly
on the fluctuations of the screening potential. 

In single-channel conductors, the asymmetry
is large [${\rm var}(\Phi)\simeq 0.2 (e^3/h\delta)^2$ for $N=2$,
not shown in Fig.~\ref{fig2}] and of the same 
order as a linear conductance fluctuation.
But as the number
of channels increases the asymmetry rapidly becomes much smaller than a 
linear conductance fluctuation or weak localization correction due to the 
smallness of the DOS fluctuations~\cite{bro97}.  
Therefore, experiments on few-channel conductors are most promising
for the detection of the asymmetry described here.  

{\em Conclusion}.---We have investigated departures from the
Onsager relations in mesoscopic
systems in the nonlinear regime.
Because of the screening potential, the weakly nonlinear conductance
is asymmetric under magnetic-field reversal.
We have determined the conditions under which
such departures are experimentally observed.
Our approach can be applied to
systems that exhibit similar phenomena such as
metallic nanowires~\cite{agr03} and molecular junctions~\cite{nit03}.

This problem was suggested to us by B. Spivak and D.M. Zumb\"uhl. 
We acknowledge P.W. Brouwer, S. Pilgram and P. Samuelsson
for helpful discussions. This work was supported by the Swiss National Science 
Foundation, the EU RTN under Contract No. HPRN-CT-2000-00144, 
and the Spanish MECD.

{\em Note added}.---During completion of this work we became aware 
of related work by B. Spivak and A. Zyuzin~\cite{zyu}
treating a diffusive cavity.

\end{document}